# Tunable light flow control in valley photonic crystal waveguide

Xiao-Dong Chen[1,†], Fu-Long Shi[1,†], Huan Liu[1], Jin-Cheng Lu[1], Wei-Min Deng[1], Jun-Yan Dai[2], Qiang Cheng[2,*], and Jian-Wen Dong[1,*]

[1]*School of Physics & State Key Laboratory of Optoelectronic Materials and Technologies, Sun Yat-sen University, Guangzhou 510275, China.*

[2]*State Key Laboratory of Millimeter Waves, Southeast University, Nanjing 210096, China.*

[†]These authors contributed equally to this work

[*]Corresponding author: qiangcheng@seu.edu.cn, dongjwen@mail.sysu.edu.cn

## ABSTRACT

The exploration of binary valley degree of freedom in topological photonic systems has inspired many intriguing optical phenomena such as photonic Hall effect, robust delay lines, and perfect out-coupling refraction. In this work, we experimentally demonstrate the tunability of light flow in a valley photonic crystal waveguide. By continuously controlling the phase difference of microwave monopolar antenna array, the flow of light can split into different directions according to the charily of phase vortex, and the splitting ratio varies smoothly from 0.9 to 0.1. Topological valley transport of edge states is also observed at photonic domain wall. Tunable edge state dispersion, i.e., from gapless valley dependent modes to gapped flat bands, is found at the photonic boundary between a valley photonic crystal waveguide and a perfect electric conductor, leading to the tunable frequency bandwidth of high transmission. Our work paves a way to the controllability and dynamic modulations of light flow in topological photonic systems.

# I. INTRODUCTION

Due to advanced fabrication, numerical modeling and characterization tools, the flexible manipulation of electromagnetic waves in a desired manner has been theoretically proposed and experimentally realized by elaborating structures and geometries of man-made materials [1-5]. Particularly, photonic crystals (PCs) are man-made periodic optical structures with a set of allowed and forbidden frequency bands, in which novel flow of light occurs [6-8]. For example, super-collimation effect is achieved in PCs with flat eigen-frequency contours [9, 10], and cloaking effect is realized in PCs with accidental Dirac cones at zone center [11, 12], and on-chip routing of spin-polarized light is implemented in glide-plane PC slabs [13, 14]. In the past few years, inspired by the discovery of topological insulators, topological photonics has attracted much attention as it provides a novel way to mold the flow of light [15-17]. Topological phases with nonzero gauge fields have been proposed and realized in different photonic systems [18-25]. Due to the macroscopic system sizes of PCs, the one-way propagating edge states are achieved in magnetic PCs by breaking time-reversal symmetry [26, 27], and robust edge states against impurities or defect without backscattering are realized in bianisotropic PCs [28, 29] or $C_6$-symmetric PCs [30-32] with strong photonic spin-orbital coupling.

Recently, the binary valley degree of freedom which characterizes the frequency extrema in the momentum space has been well explored as it has the potential to be used as an information carrier in next generation optoelectronics [33]. It leads to many intriguing phenomena such as chirality-locked beam splitting [34, 35], photonic Hall effect [36, 37], robust delay lines [38, 39], and perfect out-coupling refraction [40]. By achieving the local nonzero Berry curvature near K and K' points at the Brillouin zone corner, recent developments of valley PCs pave an alternative way to achieve silicon-on-insulator topological nanophotonic devices [41]. In this work, we experimentally demonstrate tunable light flow in topological photonic crystal waveguide by employing the valley degree of freedom. Two kinds of tunable behaviors have been shown. The light flow at the frequency of bulk states will split into two opposite directions. The splitting rate varies continuously from 0.9 to 0.1 by controlling the phase vortex of input sources. Topological valley transport of edge states is observed at photonic domain wall. Tunable edge state dispersion, i.e., from gapless valley dependent modes to gapped flat bands, is found at the photonic boundary between a valley photonic crystal waveguide and a perfect electric conductor, leading to the tunable frequency range of high transmission.

## II. VALLEY PHOTONIC CRYSTAL WAVEGUIDE

Let us start by considering a valley photonic crystal (VPC) waveguide shown in Fig. 1(a). It is constructed by two parallel metal plates at the bottom/top and a honeycomb lattice of ceramic rods ($\varepsilon$ = 8.5) in the middle. The middle ceramic rod array has the lattice constant of $a$ = 16.3 mm and the height of 14 mm. The unit cell (outlined by a dashed hexagon) consists of two rods with different diameters, i.e., rod A with the diameter of $d_A$ = 7.5 mm while rod B with the diameter of $d_B$ = 5.6 mm. The top and bottom metal plates are used to confine electromagnetic waves along the $z$-direction. By employing the zero-order transverse magnetic waveguide modes ($TM_0$ modes) which are uniform along the $z$-direction, such 3D VPC waveguide is designed to reproduce the band dispersion of TM modes of 2D VPC. Figure 1(d) shows the simulated lowest bulk band structure of $TM_0$ modes. Directional band gaps along the ΓM and ΓK directions are shaded by blue and pink rectangles, respectively. To prove the simulated directional band gaps, we measured the transmission spectra along the ΓM and ΓK directions, as shown in Figs. 1(c) and 1(e). The frequency range of measured transmission dip along the ΓM (ΓK) direction is labeled by the blue (pink) transparent rectangle, showing good agreement with the simulated directional band gap frequency range. In addition, the first bulk band has a frequency extrema of 5.46 GHz (marked by a yellow point) at the Brillouin zone corners: K' and K points. These two inequivalent *k*-points are time-reversal partners and hence the eigen-states at these two *k*-points have different field distributions. To see this, Figure 1(b) shows the phase of $E_z$ for K' valley state (upper panel) and K valley state (lower panel). Due to the uniformity of $TM_0$ modes, the $E_z$ phase do not change along the *z*-direction. For the K' valley state, the $E_z$ phase has a circular phase vortex which decreases counterclockwise by 2π around the center of the unit cell, and we denote it as left-hand circular polarization (LCP) phase vortex. In contrast, for the K valley state, the $E_z$ phase decreases clockwise by 2π around the center of the unit cell, and we denote it as right-hand circular polarization (RCP) phase vortex.

## III. TUNABLE EXCITATION OF VALLEY BULK STATES

Since the phase vortex differs from K' and K valley states, we can use the valley dependence of phase vortex to achieve the unidirectional excitation and/or tunable excitation of bulk state. The top view of experimental sample without the top metal plate is shown in Fig. 2(a). Above the bottom metal plate, there is a bulk VPC (green dash hexagon) and a surrounding piece of plexiglass (red dash hexagon) and the outmost air background (blue dash hexagon). Here, the plexiglass with the refractive index of 1.5 is used to compensate the parallel momentum and efficiently guide the bulk state outside the VPC. To construct sources with phase vortex, three monopoles (numbered as 1, 2 and 3 in the inset) are put at the center of the sample. These three monopoles are set with the same amplitude but with different initial phase (i.e., $\varphi_1$, $\varphi_2$, $\varphi_3$) by a three-port power divider and three phase shifters. By achieving different combination of ($\varphi_1$, $\varphi_2$, $\varphi_3$), the input sources with different phase vortex can be generated.

For example, when $\varphi_1$ = 240°, $\varphi_2$ = 120°, and $\varphi_3$ = 0°, the LCP phase vortex with counterclockwise decreasing phase will be generated [inset of Fig. 2(b)]. By putting such source with LCP phase vortex at the center of VPC, only K' valley state can be excited at the working frequency of 5.46 GHz [Fig. 2(b)]. At each boundary, the K' valley states are partially reflected back into the VPC and partially refracted into the surrounding plexiglass. The refracted light beams enhance at the left, upper-right, and lower-right corners of the whole sample. In contrast, when the input source carries another configuration of $\varphi_1$ = 0°, $\varphi_2$ = 120°, $\varphi_3$ = 240° (i.e., RCP phase vortex source), only K valley state will be excited, and the refracted light beams will switch to enhance at the right, upper-left, and lower-left corners of the whole sample. To quantitatively demonstrate such chiral-source dependent splitting behaviors, we measured the magnitude of $E_z$ fields at the left and right exit, as outlined by white frames. Meanwhile, we have continuously changed the phase vortex source from LCP to linear polarization and RCP, in order to investigate the splitting behaviors. In the experiment, we keep $\varphi_2$ as 120° but decrease $\varphi_1$ from 240° to 0° and increase $\varphi_3$ from 0° to 240°. The phase vortex source and VPC waveguide are put on the bottom metal plate which is stationary during the measurement. On the contrary, the top metal plate is mounted on a two dimensional motorized translation stage (LINBOU NFS03). A hole is drilled in the top plate, and a monopole antenna is inserted to record the electric fields which will be collected by the vector network analyzer (Agilent E5071C). All the measured results near the left and right exit are summarized in Fig. 2(d). The corresponding simulation results are also shown in Fig. 2(c). Here, we list the near-field distributions of five different phase configurations. When the excited source has LCP phase vortex, i.e., $\varphi_1$ = 240°, $\varphi_2$ = 120°, $\varphi_3$ = 0° of the first case, the $E_z$ fields at the left exit are much larger than those at the right exit, confirming the unidirectional excitation of photonic states in the topological photonic crystal by the chiral source. By decreasing $\varphi_1$ while increasing $\varphi_3$ simultaneously, we measured the output near-field distributions for the other four cases. One can see that both simulated and measured magnitude at the left exit gradually become smaller while those at the right exit become larger when the input source changes. To characterize the tunable output field distribution, we focus on $E_z$ at 20 mm away from boundary between plexiglass and air [marked by white dots in the third case of Figs. 2(c) and 2(d)]. We define $r_{\mathrm{L}} = |E_z^{\mathrm{L}}| / \left( |E_z^{\mathrm{L}}| + |E_z^{\mathrm{R}}| \right)$ and $r_{\mathrm{R}} = |E_z^{\mathrm{R}}| / \left( |E_z^{\mathrm{L}}| + |E_z^{\mathrm{R}}| \right)$ to show the splitting ratio of the field magnitude at the left and right exits. As shown in Fig. 2(e), the splitting ratio (red curve) at the left exit $r_{\mathrm{L}}$ is 0.9 for LCP phase vortex source. It gradually becomes smaller along with the decreasing $\varphi_1$ (meanwhile increasing $\varphi_3$), and at last reaches 0.1 when the source has RCP phase vortex. The measured experimental splitting ratio $r_{\mathrm{L}}$ and $r_{\mathrm{R}}$ are good agreement with the simulated results. Note that there is little mismatch when $\varphi_1$ ranges from 160° to 80° [dark grey in Figs. 2(e) and 2(f)]. This is because the electric fields have small magnitude, e.g., the third column data in Figs. 2(c) and 2(d), leading to a relatively large signal-noise error. For the practical application, one may focus on the ranges of $\varphi_1$ from 240° to 170° and 70° to 0° [light grey in Figs. 2(e) and 2(f)] to achieve the splitting ratio of electric fields from 0.9 to 0.1.

## IV. ROBUST TRANSPORT OF VALLEY EDGE STATES

Another important property of the VPC waveguide is broadband robust transport of edge states. Figure 3(a) shows the top view schematic of the photonic domain wall. The VPC waveguide presented in Fig. 1(a) [denoted as VPC1 for short] and its inverted one [denoted as VPC2 for short] locate above and below the domain wall, respectively. Previous theoretical results indicate that the topological indices of VPC1 (VPC2) are $C_{\mathrm{K'}}$ = +1/2 and $C_{\mathrm{K}}$ = -1/2 ($C_{\mathrm{K'}}$ = -1/2 and $C_{\mathrm{K}}$ = +1/2) [42]. Across the boundary, the topological indices difference is -1 and +1 at K' and K valleys, respectively. According to the bulk edge correspondence, there will be one edge state with a negative (positive) group velocity at the K' (K) valley, which is confirmed by the numerical edge dispersion shown in Fig. 3(b). When the inter-valley scattering is prohibited, robust transport of these valley-dependent edge states can be observed. To see this, we construct two different photonic boundaries, i.e., the flat channel and Z-shape bend [Figs. 3(d) and 3(e)]. Here the top metal plate is removed to show the boundary morphology inside (outlined by blue and red dash lines). Figure 3(c) shows the measured transmission spectra for both flat channel (blue line) and Z-shape bend (red line). Within the frequency range of edge states [shaded by a blue transparent rectangle], high transmission is preserved even when the excited edge states go along the Z-shape bend with two sharp corners. To image the robust transmission directly, we also scanned the transmitted $E_z$ fields [Figs. 3(f) and 3(g)]. In the experiment, the edge states are excited by a monopole locating at the left entrance of two boundaries. The top metal plate is mounted on a two dimensional (*xy* plane) motorized translation stage. A scanning antenna is inserted into the top plate to record $E_z$, and the signals are collected by the vector network analyzer (Agilent E5071C). From the $E_z$ magnitude distributions shown in Figs. 3(f) and 3(g), we can directly see the edge states pass along the Z-shape bend without scattering and the robust transport of valley edge states is demonstrated.

The recent reported valley dependent edge states are achieved at the domain wall between two VPCs with distinct topology [37, 38]. To miniaturize the boundary size by half, next we show that such valley dependent edge states can also be obtained at the boundary between only one VPC and a perfect electric conductor (PEC) [left schematic of Fig. 4(a)]. The distance between the PEC and the nearest rod (highlighted in red) is set to be 4 mm. To obtain edge states whose frequency range covering the whole complete band gap, we tune the diameter of the nearest rods to be $d$ = 5.5 mm. Figure 4(b) shows the gapless edge dispersion of this boundary. The edge states with a positive group velocity appear at the K' valley while the edge states have a negative group velocity at the K valley. In searching for the origin of these valley dependent edge states, we notice that similar gapless edge states also exist at the topological domain wall of two VPCs [right schematic of Fig. 4(a)]. Considering the odd modes of domain wall, the electric fields are required to be perpendicular to the mirror symmetric plane (black dash line), as if an effective PEC boundary is there. Therefore, solving the odd modes of topological domain wall of two VPCs [right schematic of Fig. 4(a)] is equivalent to solving the eigenmodes of the edge states of PEC-capped VPC [left schematic of Fig. 4(a)]. Note that the even modes of topological domain wall of two VPCs can be mimicked by the VPC with a perfect magnetic conductor [which is typically realized with artificial resonant structure and hamper the experimental implementation]. However, comparing to the domain wall, the simplified PEC-capped VPC waveguide can miniaturize the sample size by half. The edge states can evolve continuously from the gapless valley dependent modes to the flat band by tuning the magnitude of the photonic potential near the boundary, when the topological property of bulk VPC is unchanged. To see this, we increase the photonic potential by enlarging the diameter of the nearest rods from 5.5 mm to 6.5 mm [Fig. 4(c)] and 7.5 mm [Fig. 4(d)]. When the diameter of nearest rods becomes larger, more energy of edge states locates at the ceramic rods with high permittivity, and the edge dispersions are shifted to lower frequency, achieving the evolution from gapless edge states to flat dispersion bands. In addition, the frequency range of edge states is easily tuned by changing the diameter of ceramic rods, serving as a platform for realizing tunable frequency bandwidth with high transmission. We also simulated the transmission spectra of the tunable edge states of PEC-capped VPC, as shown in Fig. 4(e). The source is incident from the left and the transmission is recorded at the right end of boundary. Figure 4(f) illustrates high transmission in the whole complete band gap frequency range when $d$ = 5.5 mm [red curve]. When the rod diameter is changed to be 6.5 mm, transmission near 6 GHz drops [grey solid arrow] while it keeps unchanged near 5.7 GHz. When the rod diameter is further enlarged to be 7.5 mm, transmission at 5.7 GHz drops [grey dash arrow]. Therefore, tunable high transmission of edge states can be achieved by increasing the rod diameter. Note that similar tunable high transmission can be also realized by changing the distance between rods and PEC.

## V. CONCLUSION

In conclusion, we design and fabricate a VPC waveguide by sandwiching a honeycomb lattice of ceramic rods between two parallel metal plates. Using the phase vortex of valley bulk states, tunable flow of light is experimentally achieved in such VPC waveguide. The flow of light splits into different directions and the splitting ratio can vary from 0.9 to 0.1 continuously. Furthermore, at the topological domain wall between two distinct VPC waveguides, robust transport of edge states is observed. In addition, we demonstrate tunable edge state dispersion, i.e., from gapless valley dependent modes to gapped flat bands, at the photonic boundary between a VPC waveguide and a perfect electric conductor, leading to the feasibility to manipulate frequency range of high transmission.

## ACKNOWLEDGEMENTS

This work is supported by Natural Science Foundation of China (11704422, 61775243, 11522437), and Fundamental Research Funds for the Central Universities (No.17lgpy19).

## FIGURES AND FIGURE CAPTIONS

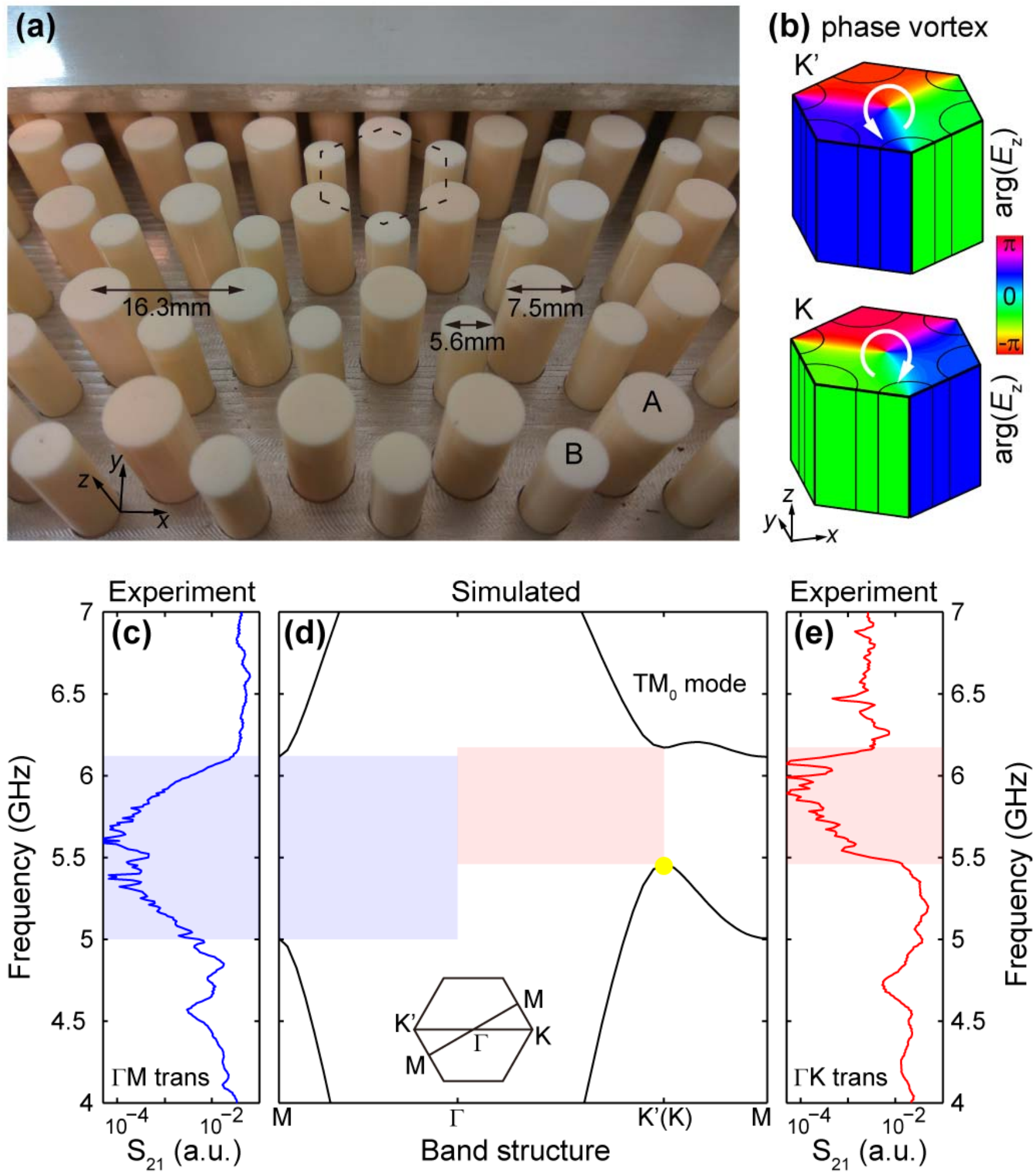

FIG. 1. Experimental sample and bulk band structure of valley photonic crystal waveguide. (a) Photo of the valley photonic crystal waveguide. It has a hexagonal lattice of ceramic rods with the lattice constant of $a$ = 16.3 mm. Within a unit cell (black dash hexagon), there are two kinds of rods with different diameters of 7.5 mm (marked as rod A) and 5.6 mm (rod B), respectively. The rod array is then covered by two parallel metal plates. (b) Valley-contrasting chiral phase distributions of $E_z$, i.e., arg($E_z$), at the frequency of 5.46 GHz for bulk states at K' and K point [marked by a yellow point in (d)]. (c-e) Experiment transmission spectra (c,e) and calculated bulk band structures (d) of zero-order transverse magnetic waveguide modes ($TM_0$ modes). The directional band gaps along the ΓM and ΓK directions are labeled by blue and pink transparent rectangles, respectively. Inset of (d) shows the Brillouin zone with high symmetry $k$-points. The measured transmission spectra along the (c) ΓM and (e) ΓK directions are used to confirm the frequency range of calculated directional band gaps.

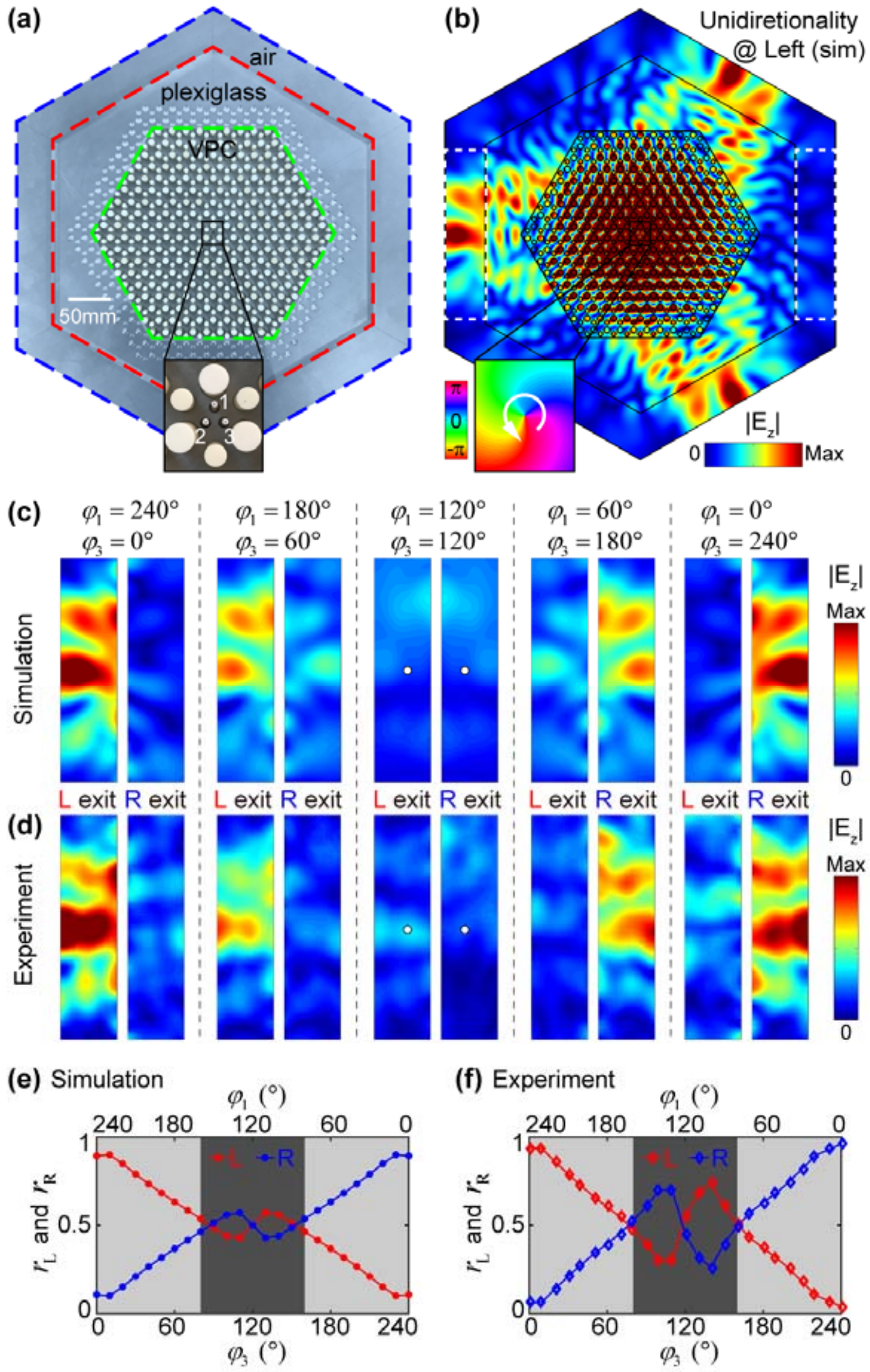

FIG. 2. Tunable excitation of bulk states. (a) Photo of the bulk valley photonic crystal (VPC) waveguide with the removal of the top metal plate to show inside. Between the top and bottom metal plates, there is a honeycomb lattice of ceramic rods (green dash hexagon), a piece of plexiglass with the refractive index of 1.5 (red dash hexagon) and the outmost air background (blue dash hexagon). Inset: A self-made source is constructed by three monopoles (labeled as 1, 2, and 3) with different initial phases (i.e., $\varphi_1$, $\varphi_2$, $\varphi_3$). (b) Simulated magnitude of $E_z$ when the VPC waveguide is excited by source with LCP phase vortex at the frequency of 5.46 GHz. Two white frames outline the regions where $E_z$ are measured. Inset: The phase vortex with counterclockwise decreasing phases is generated when $\varphi_1 = 240°$, $\varphi_2 = 120°$, and $\varphi_3 = 0°$ is input. (c) Simulated and (d) measured magnitude of $E_z$ at left exit and right exit of the bulk sample when $\varphi_2$ is kept as 120° but $\varphi_1$ and $\varphi_3$ are changed. (e) Simulated and (f) measured splitting ratio of electric fields at the left and right exits. Here, $E_z$ at 20 mm away from the boundary (marked by white dots in (c) and (d)) are chosen to be the representative electric fields. The range of $\varphi_1$ from 160° to 80° (shaded in dark grey) is out of the discussed scope as the corresponding electric fields at the left and right exits are too small. The splitting ratio from 0.9 to 0.1 can be achieved when $\varphi_1$ ranges from 240° to 170° and 70° to 0°.

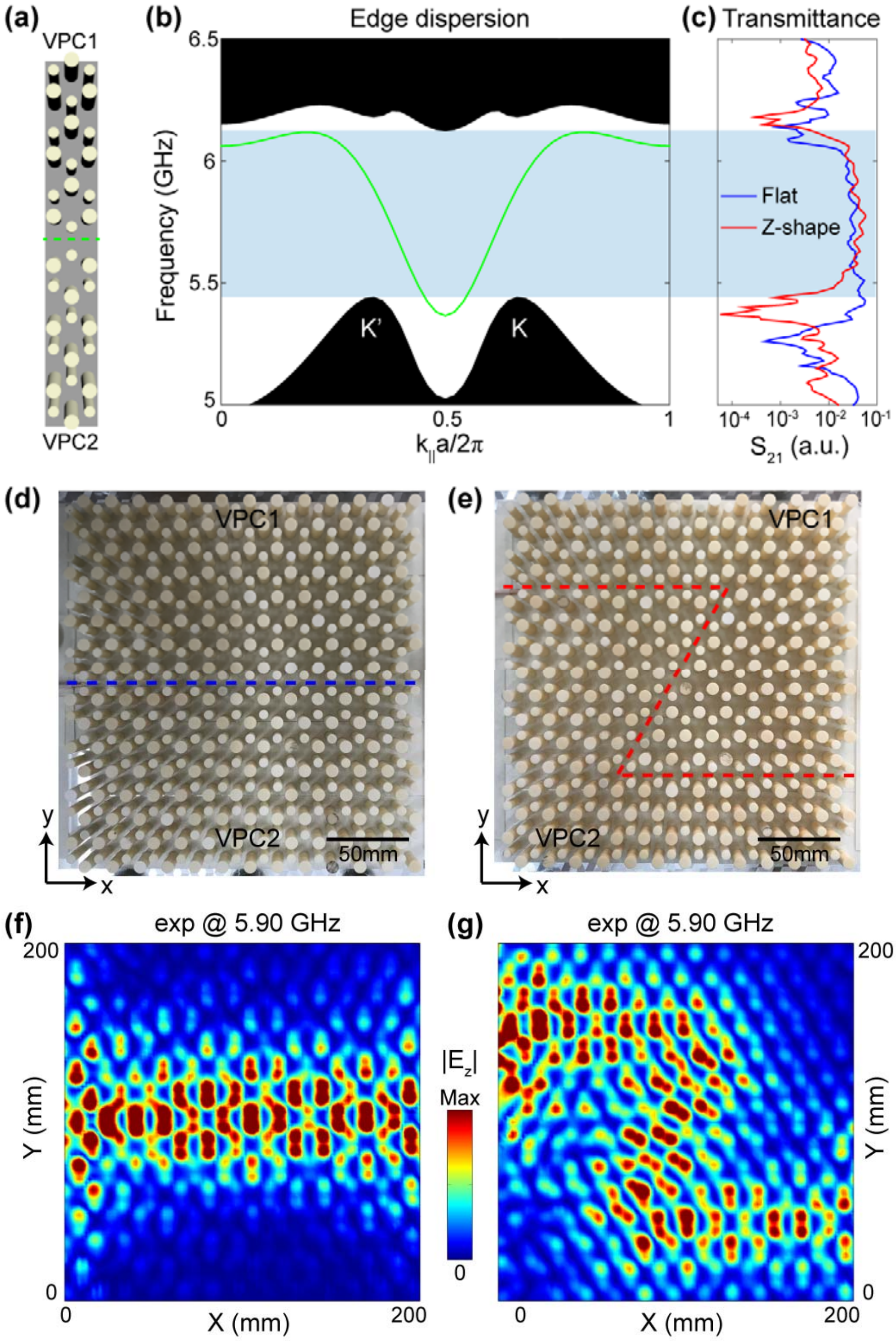

FIG. 3. Observation of broadband robust transport of valley edge states. (a) Schematic of photonic boundary between two topologically-distinct valley photonic crystals (VPC1 and VPC2). (b) Edge dispersion corresponding to the boundary in (a). The green line represents the edge states and the black region corresponds to the projection bulk bands. (c) Measured transmission spectra of the flat channel (blue curve) and the Z-shape bend (red curve). The complete band gap is labeled in both (b) and (c) by a blue transparent rectangle. (d, e) Photos of (d) the flat channel and (e) the Z-shape bend between VPC1 and VPC2. The interface is outlined by the blue and red dash lines, respectively. (f, g) Measured magnitude of $E_z$ for (f) the flat channel and (g) the Z-shape bend at the frequency of 5.9 GHz.

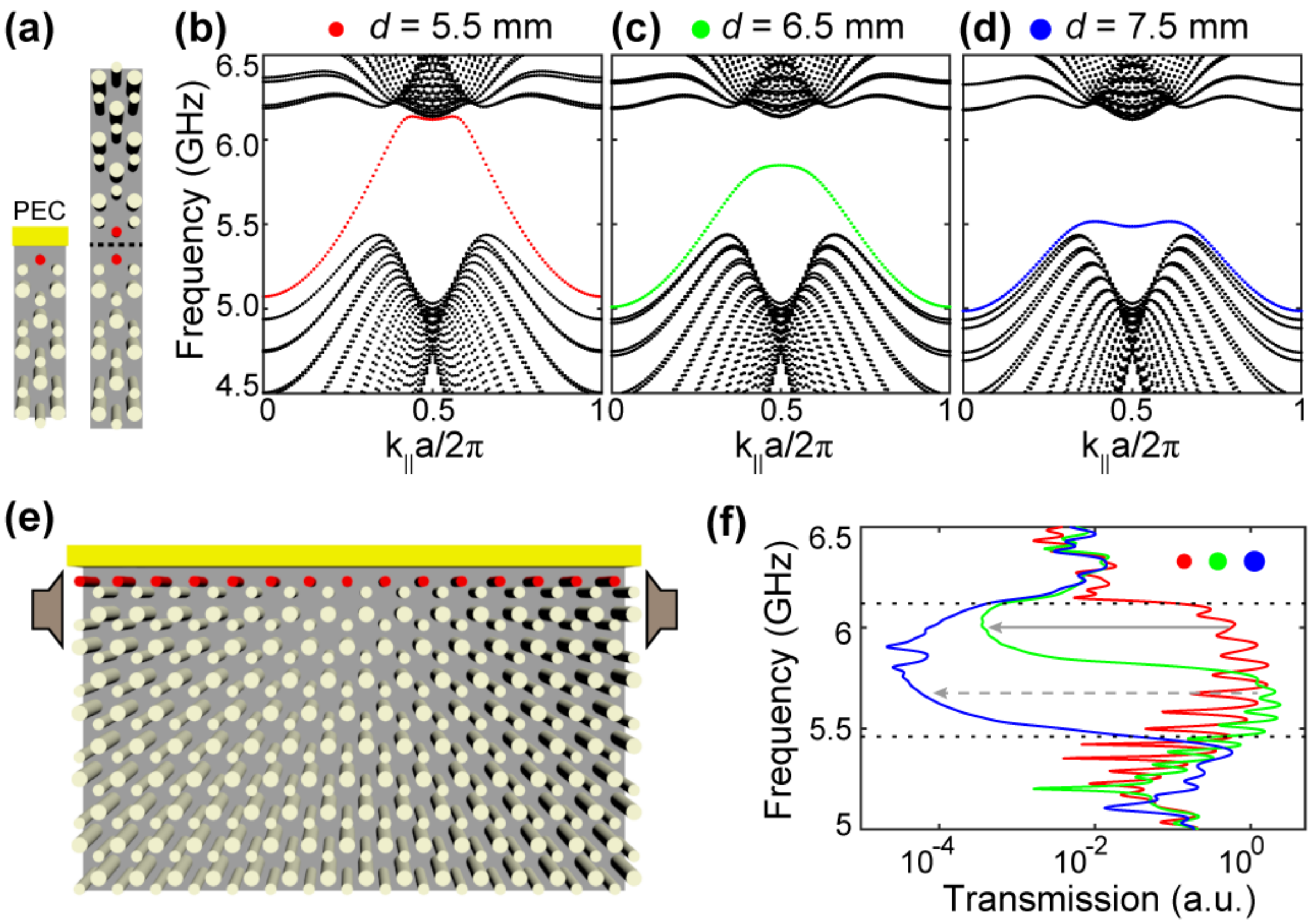

FIG. 4. Tunable edge dispersion from gapless valley dependent modes to gapped flat band. (a) Schematics of the boundary between a VPC waveguide and a PEC (left) and the topological domain wall between two VPC waveguides (right). (b-d) Edge state dispersion for photonic boundary with nearest ceramic rods having the diameter of (b) $d = 5.5$ mm, (c) $d = 6.5$ mm, and (d) $d = 7.5$ mm. Here, the distance between the center of the nearest rod and PEC is 4 mm. The evolution from gapless valley dependent modes to gapped flat band are observed. (e) Schematic of the photonic boundary for tunable high transmission. Source is incident from the left, and the transmission is recorded at the right. (f) Transmission spectra of tunable frequency range of high transmission. Solid grey arrow (dash grey arrow) marks that the transmission at 6 GHz (5.7 GHz) drops when rod diameter is changed to be $d = 6.5$ mm ($d = 7.5$ mm). Two black dash lines mark the frequency range of the complete band gap.